\documentclass[%
 reprint,
 amsmath,amssymb,
 aps,
]{revtex4-2}
\usepackage{graphicx}
\usepackage{subcaption}
\usepackage{graphicx}
\usepackage{dcolumn}
\usepackage{bm}
\usepackage{subcaption}

\begin{document}

\preprint{APS/123-QED}

\title{Unveiling the Mechanisms of Electron Energy Spectrum Evolution}

\author{Xu-Lin Dong}
 \affiliation{College of Physics, Hebei Normal University, No. 20 Road East 2nd Ring South, Shijiazhuang, 050024 Hebei, China}
\affiliation{
Key Laboratory of Particle Astrophysics, Institute of High Energy Physics, Chinese Academy of Sciences, No. 19 B Yuquan Road, Shijingshan District, Beijing, 100049 Beijing, China}

\author{Shu-Wei Ma}
\affiliation{College of Physics, Hebei Normal University, No. 20 Road East 2nd Ring South, Shijiazhuang, 050024 Hebei, China }
\affiliation{Key Laboratory of Particle Astrophysics, Institute of High Energy Physics,Chinese Academy of Sciences, No. 19 B Yuquan Road, Shijingshan District, Beijing, 100049 Beijing, China}

\author{Yi-Qing Guo}\email{guoyq@ihep.ac.cn}
\affiliation{Key Laboratory of Particle Astrophysics, Institute of High Energy Physics, 
Chinese Academy of Sciences, Beijing 100049, China}
\affiliation{University of Chinese Academy of Sciences, Beijing 100049, China}
\affiliation{TIANFU Cosmic Ray Research Center, Chengdu, 610213 Sichuan, China}

\author{Shu-Wang Cui}\email{cuisw@hebtu.edu.cn}
\affiliation{College of Physics, Hebei Normal University, No. 20 Road East 2nd Ring South, Shijiazhuang, 050024 Hebei, China}

\date{\today}

\begin{abstract}

The electron spectrum exhibits a complex structure and has controversially proposed origins. This work reproduce the evolution of the electron spectrum based on a spatially dependent propagation (SDP) model.  The key point is that our SPD model features two diffusion regions leading to two diffusion timescales, competing with the cooling timescale. This results in a three-segment power-law electron spectrum: (1) The spectrum below tens of GeV is primarily influenced by cooling effects from distant sources. (2) The spectrum dominated by diffusion effects from nearby sources from tens of GeV to TeV. (3) The spectrum above TeV, which is predominantly governed by cooling effects from nearby sources. This evolution is unique to the SDP model, and we offer a comprehensive and clear depiction of electron evolution under a single propagation scenario for the first time.

\end{abstract}

\maketitle


\section{INTRODUCTION}

High-energy electron spectra serve as an excellent probe for nearby sources of cosmic rays (CRs). The electron propagation time scale can be expressed as 
$\tau_{dif}=L^2/D_{xx}$, where $L$ represents the diffusion halo radius and $D_{xx}$ denotes the diffusion coefficient. The electron energy loss time scale can be approximated as \cite{blumenthal1970bremsstrahlung,cholis2022utilizing}
\begin{equation}
\tau_{loss}(E_{init})\simeq20\times(\frac{E_{init}}{10GeV})^{-1}Myr
\end{equation}
 where $E_{init}$ represents the initial energy of the electron. When $\tau_{loss}<\tau_{dif}$ is satisfied, energy loss becomes dominant, thereby obscuring the contributions from distant sources to the high-energy component, while the number of nearby sources is minimal. Therefore, high-energy electron spectra serve as an excellent probe for nearby sources. To conduct a thorough investigation using this probe, precise measurements are essential.

Experiments such as AMS-02, DAMPE, and HESS have provided comprehensive and precise measurements of the electron spectrum\cite{aguilar2021alpha,adriani2023direct,dampe2017direct,abdollahi2017cosmic,kerszberg2017cosmic,aharonian2024high,archer2018measurement}. The electron spectrum shows hardening around ${\sim}40\mathrm{GeV}$\cite{aguilar2021alpha,adriani2023direct,dampe2017direct} and experiences cutoff at ${\sim}\mathrm{TeV}$\cite{kerszberg2017cosmic,aharonian2024high}. These reveal the intricate three-segment power-law structure of the electron spectrum, prompting theoretical physicists to utilize a range of theoretical models to elucidate its origins.

 Currently, the interpretations of the electron spectrum primarily involve neighboring source models such as pulsar wind nebulae\cite{zhang2021possible,evoli2021galactic,hooper2017hawc,bykov2019gev,evoli2020signature} and supernova remnants\cite{kobayashi2004most,di2014interpretation,fang2018explanation}, the K-N effect\cite{fang2021klein,evoli2020signature}, and dark matter particle annihilation\cite{liu2018tev,huang2018origins,feng2020interpretation}. However, the crucial issue is that these models fail to dynamically portray the mechanisms underlying the evolution of the electron spectrum. Therefore, it is essential to provide a comprehensive and clear depiction of electron evolution.

In this work, we trace the origins of the evolution of the electron spectrum using SDP model.The organization of this paper is as follows: Sec.II presents CR SDP model, followed by the results in Sec.III. Finally, Sec.IV provides a summary.

\begin{figure*}
\includegraphics[width=0.9\textwidth]{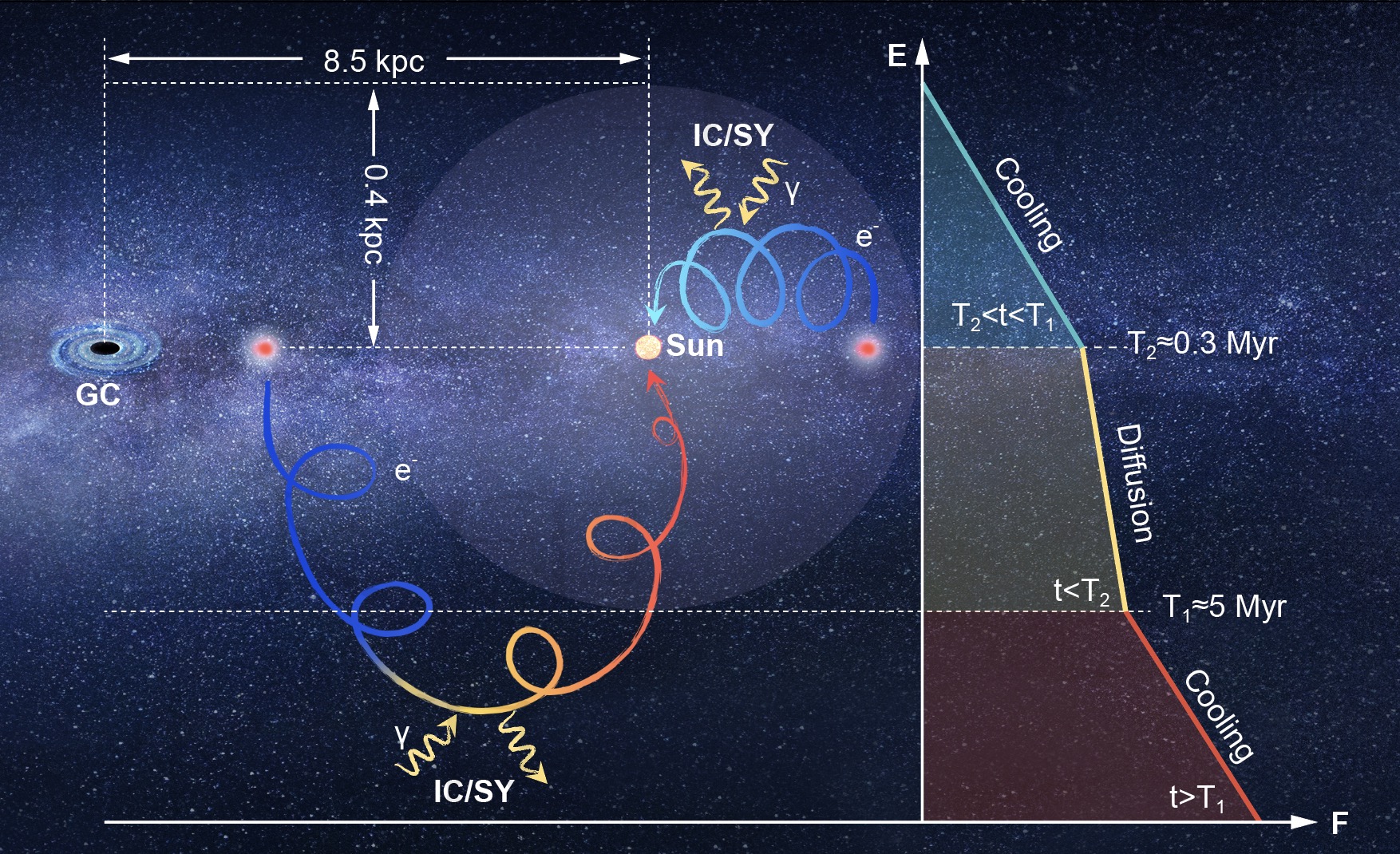}
\caption{\label{fig:wide}Schematic illustration of the physical mechanisms underlying the three segments of the power-law spectrum of CR electrons. Electrons from distant sources propagate through the outer zone, where they are dominated by cooling effects and concentrated in the low-energy region, forming the first segment of the energy spectrum. In contrast, electrons contributed by nearby sources are concentrated in the high-energy region and are classified into two parts based on their energy, with one part dominated by diffusion and the other by cooling, leading to the second and third segments of the electron energy spectrum, respectively. \texttt{}}
\end{figure*}

\section{MODEL AND METHODOLOGY}

To describe the propagation process of CRs, we utilize the SDP model\cite{tomassetti2012origin,gaggero2015gamma,guo2015spatial,liu2018revisiting,yao2024common,abeysekara2017extended} in this work. As shown in Figure . 1, the presence of a slow diffusion region around the source, combined with the primary distribution of sources on the galactic plane, necessitates the division of the CR propagation space into inner and outer zones. The values of $D_{xx}$ and $L$ differ between the inner and outer zones, leading to two distinct diffusion timescales, $\tau_{IH}$ and $\tau_{OH}$, within the SDP model. Previous work has demonstrated that particles contributed by distant sources (Comp-A) propagate through the outer zone, resulting in longer propagation times and lower energies, while particles contributed by nearby sources (Comp-B) propagate through the inner zone, leading to shorter propagation times and higher energies. Consequently, for electrons from distant sources, $\tau_{loss}<\tau_{OH}$, and their propagation process is predominantly governed by cooling effects, thereby forming the first segment of the electron spectrum. Electrons from nearby sources can be categorized into two groups based on energy: those with energies $\tau_{loss}>\tau_{IH}$, whose propagation process is dominated by diffusion effects, constitute the second segment of the electron spectrum; and those with energies $\tau_{loss}<\tau_{IH}$, whose propagation process is governed by cooling effects, form the third segment of the electron spectrum. Therefore, this group of electrons should exhibit a distinct structural transition from diffusion dominance to cooling dominance. The combined contributions of these three components lead to the complex structure of the electron spectrum. Based on the location of the inflection point in the electron spectrum, it can be inferred that $\tau_{OH}\approx5Myr$ and $\tau_{IH}\approx0.3Myr$.

\begin{table*}
\caption{\label{tab:table3}Propagation model parameters and spectral injection parameters.}
\begin{ruledtabular}
\begin{tabular}{ccccccccccccc}
 Model&$D_0{}^{\mathrm{a}}[\mathrm{cm}^{-2}\mathrm{s}^{-1}]$ &$\delta_0$&$\quad N_m$&$\quad\xi$&$\quad n$&$\quad v_A[\mathrm{km~s}^{-1}]$&$\quad z_0[\mathrm{kpc}]$&$\text{Normalization}[\mathrm{GeV}^{-1}\mathrm{m}^{-2}\mathrm{s}^{-1}\mathrm{sr}^{-1}]$&$\quad\nu_{1}^{\mathrm{a}}$&$\mathcal{R}_{\mathrm{br}}[\mathrm{GV}]$&$\quad\nu_{2}^{\mathrm{a}}$ \\ \hline
 SDP-1&$1\times10^{29}$&0.8&0.6&0.082&4.0&6&5&0.268&1.85&7.8&2.68 \\
 SDP-2&$1\times10^{29}$&0.8&0.6&0.082&4.0&6&5&0.268&1.85&7.8&2.72\\
 TRO&$1\times10^{29}$&0.33&0.6&0.0&4.0&6&5&0.267&1.85&7&2.8\\
\end{tabular}
\end{ruledtabular}
\end{table*}

\begin{figure*}[t]
    \centering
    \begin{subfigure}{0.48\textwidth}
        \centering
        \includegraphics[width=\textwidth]{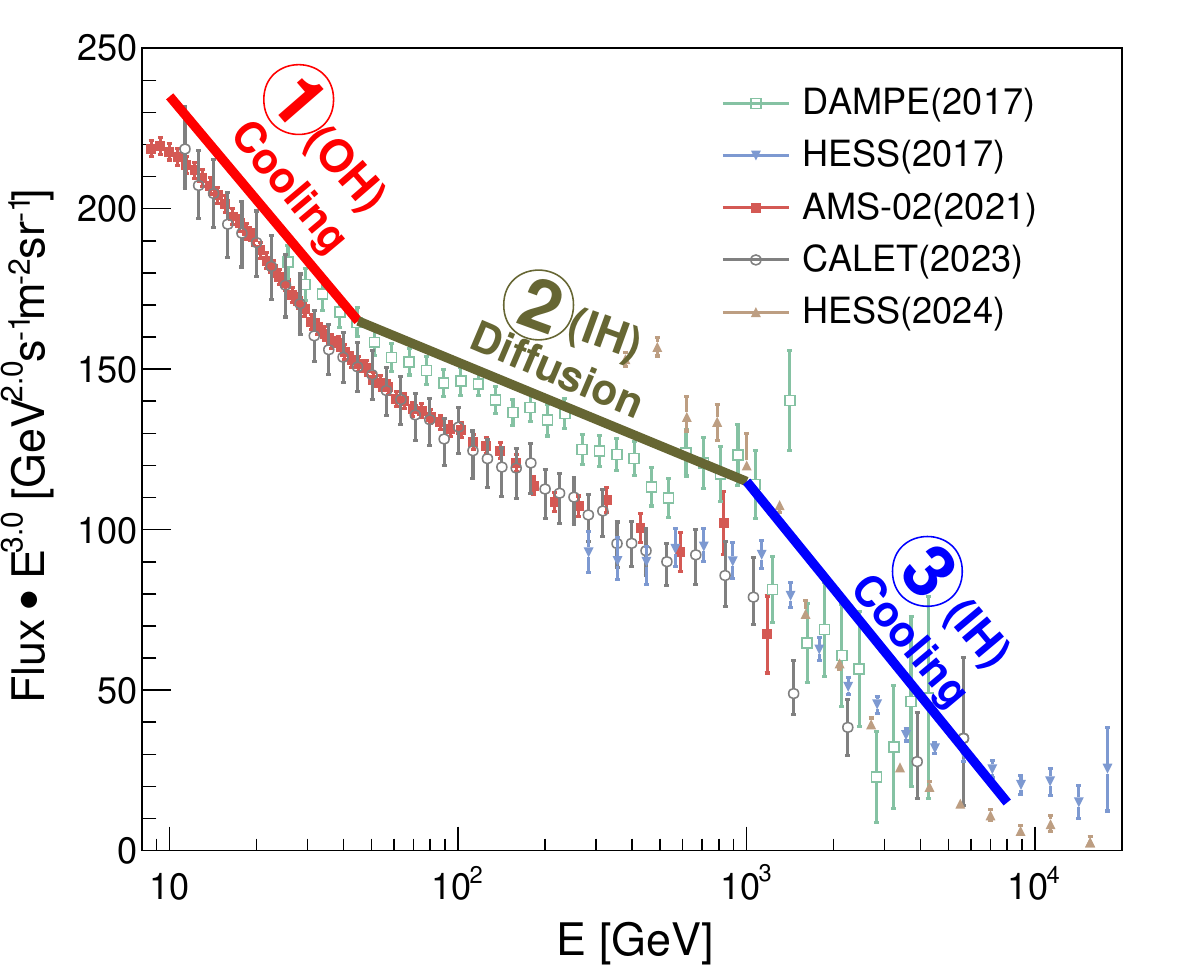}
    \end{subfigure}
    \hfill
    \begin{subfigure}{0.48\textwidth}
        \centering
        \includegraphics[width=\textwidth]{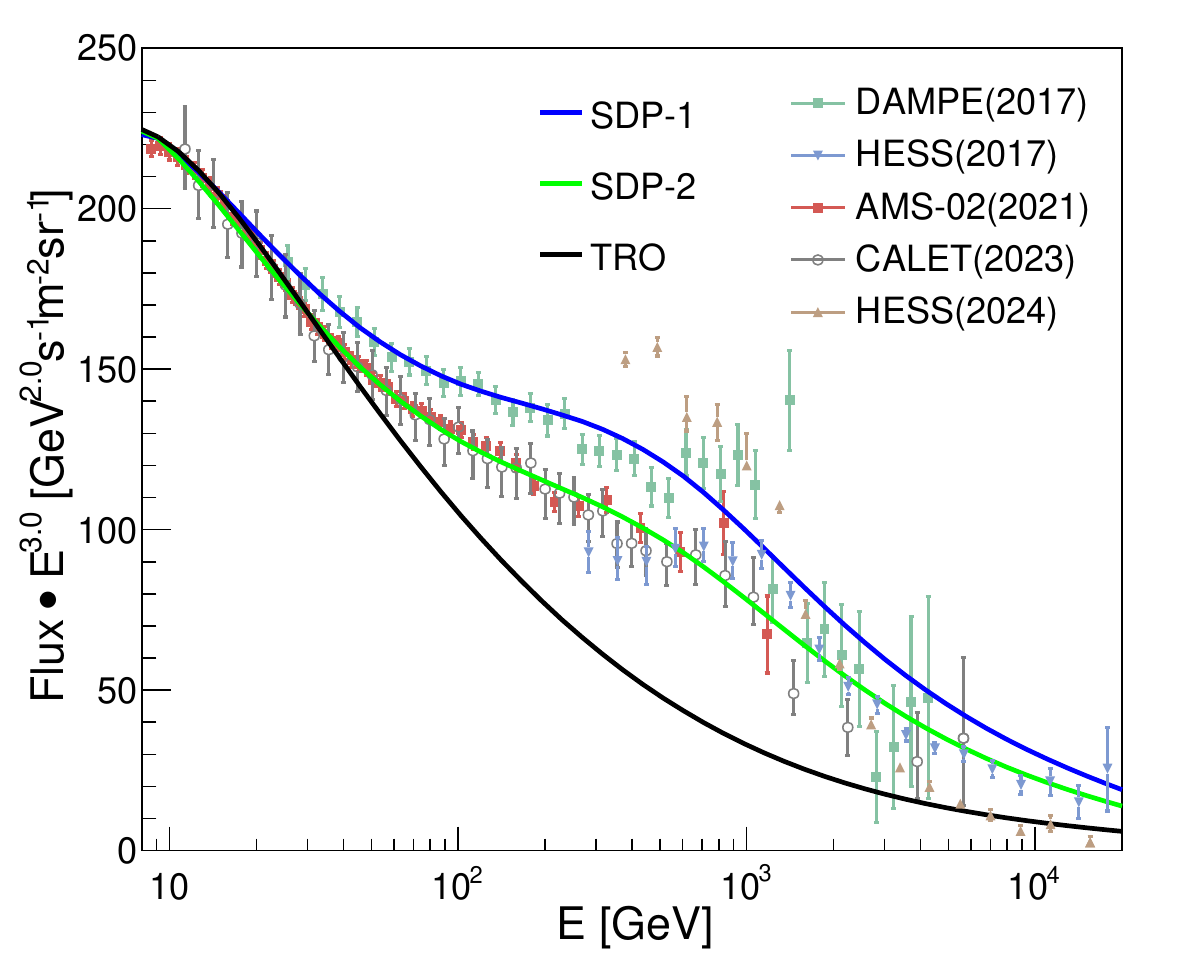}
    \end{subfigure}
    \caption{Left panel: Electron spectrum observed by the AMS-02\cite{aguilar2021alpha}, DAMPE\cite{dampe2017direct}, HESS\cite{kerszberg2017cosmic,aharonian2024high}, and CALET\cite{adriani2023direct} experiments.  Right panel: Measurement results from each experiment compared with the calculational results from the SDP model and the traditional model. The black line represents the results from the traditional model, while the blue and green lines correspond to the calculational results from the SDP model for the AMS-02 and DAMPE experiment results, respectively.}
    \label{fig:twosub}   
\end{figure*}

In the following a quantitative description of the CR propagation process is provided. The propagation of CRs can be represented mathematically by the following partial differential equation\cite{maurin2002galactic}
\begin{equation}
\begin{split}
\frac{\partial\Psi(\vec{r},p,t)}{\partial t} = Q(\vec{r},p,t)+\vec{\nabla}\cdot(D_{xx}\vec{\nabla}\Psi-\vec{V}_{c}\Psi)+ \\
\frac{\partial}{\partial p}[p^{2}D_{pp}\frac{\partial}{\partial p}\frac{\Psi}{p^{2}}]-\frac{\partial}{\partial p}[\dot{p}\Psi-\frac{p}{3}(\vec{\nabla}\cdot\vec{V}_{c})]-\frac{\psi}{\tau_{f}}-\frac{\psi}{\tau_{r}},
\end{split}
\end{equation}

where $Q(\vec{r},p,t)$ describes the distribution of sources,$\vec{V}_{c}$ represents the convection velocity, $D_{xx}$ denotes the spatial diffusion coefficient, $D_{pp}$ is the momentum diffusion coefficient accounting for the reacceleration process, $\dot{p}$,$\tau_{f}$ and $\tau_{r}$ are the energy loss rate, the fragmentation and radioactive decaying timescaleds, respectively. 

Since the diffusion speed of CRs in the inner zone is lower than that in the outer zone and is dependent on the distribution of sources, the diffusion coefficient of the model is expressed as:
\begin{equation}D_{xx}(r,z,R)=D_{0}F(r,z)\beta^{\eta}(\frac{R}{R_{0}})^{\delta(r,z)},\end{equation}
where the function $F(r,z)$ is defined as:
\begin{equation}F(r,z)=\begin{cases} g(r,z)+[1-g(r,z)]\left(\frac{z}{\xi z_0}\right)^n,&|z|\le\xi z_0\\ 1,&|z|>\xi z_0\end{cases},\end{equation}
With $g(r,z)=N_m/[1+f(r,z)]$. Here the diffusion coefficient of the inner zone is anticorrelated with the source distribution $f(r,z)$ , given by
\begin{equation}f(r,z)=\left(\frac{r}{r_\odot}\right)^{1.25}\exp\biggl[-\frac{3.87(r-r_\odot)}{r_\odot}\biggr]\exp\biggl(-\frac{|z|}{z_s}\biggr),\end{equation} 
where $r_{\odot}=8.5\text{ kpc}$ and $z_s=0.2\text{ kpc}$ . For the outer zone, the diffusion coefficient remains constant when varying spatial locations.In contrast, the diffusion coefficient in the traditional model remains constant throughout the entire space. 
In order to investigate the relationship between the electron spectra at different energy ranges and the locations of their sources, we categorize $f(r,z)$ into two parts. Using the position of the Sun as the central reference point, we divide the sources into nearby and distant sources based on their distance $R$ from the Sun.

The source spectrum of CRs is assumed to be a broken power law in rigidity. To solve the transport equation, we employ the numerical package DRAGON\cite{evoli2008cosmic} . The forcefield approximation is incorporated to account for solar modulation effects\cite{gleeson1968solar}. The key parameters pertaining to CR propagation are summarized in Table I.

\section{RESULTS}
\begin{figure*}[t]
    \centering
    \begin{subfigure}{0.48\textwidth}
        \centering
        \includegraphics[width=\textwidth]{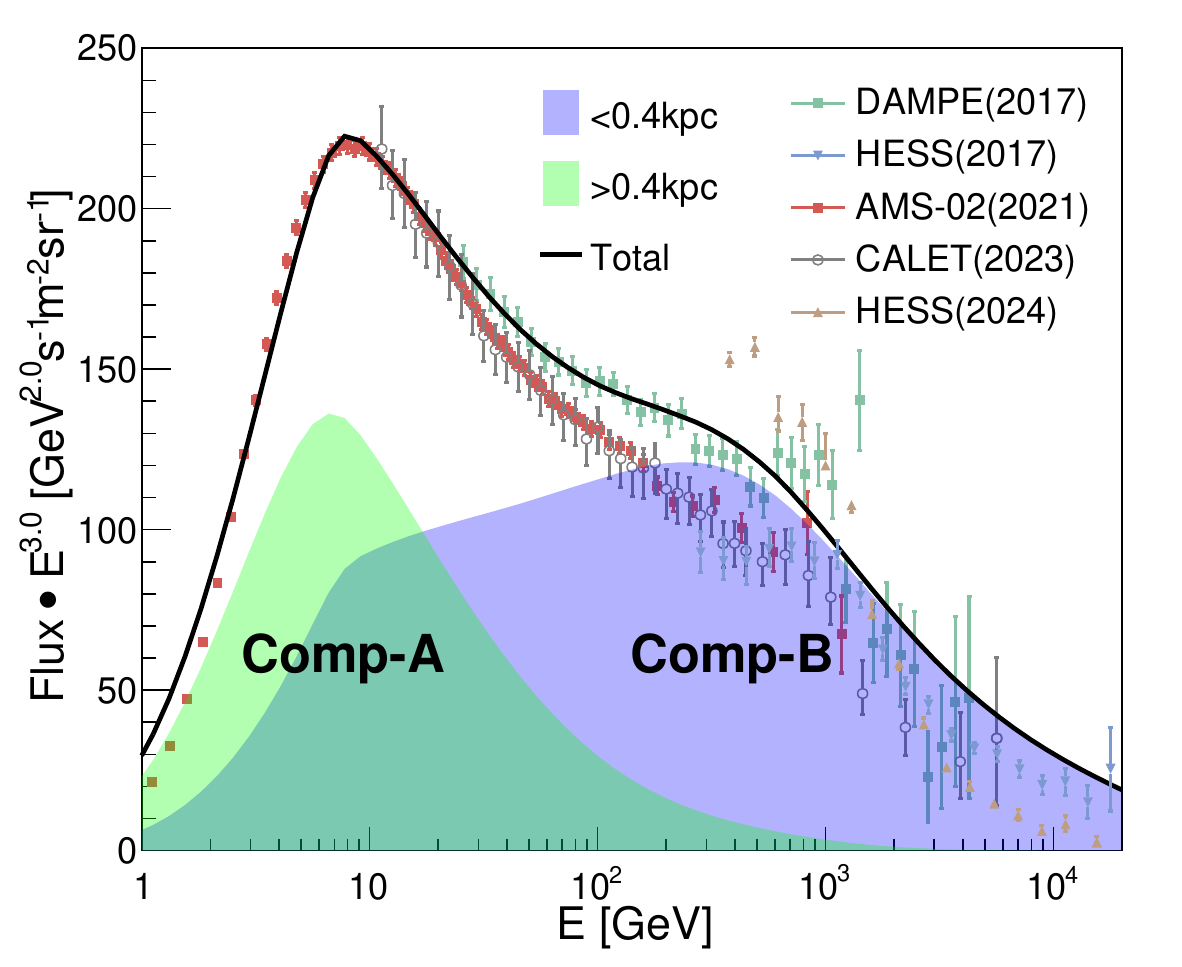}
    \end{subfigure}
    \hfill
    \begin{subfigure}{0.48\textwidth}
        \centering
        \includegraphics[width=\textwidth]{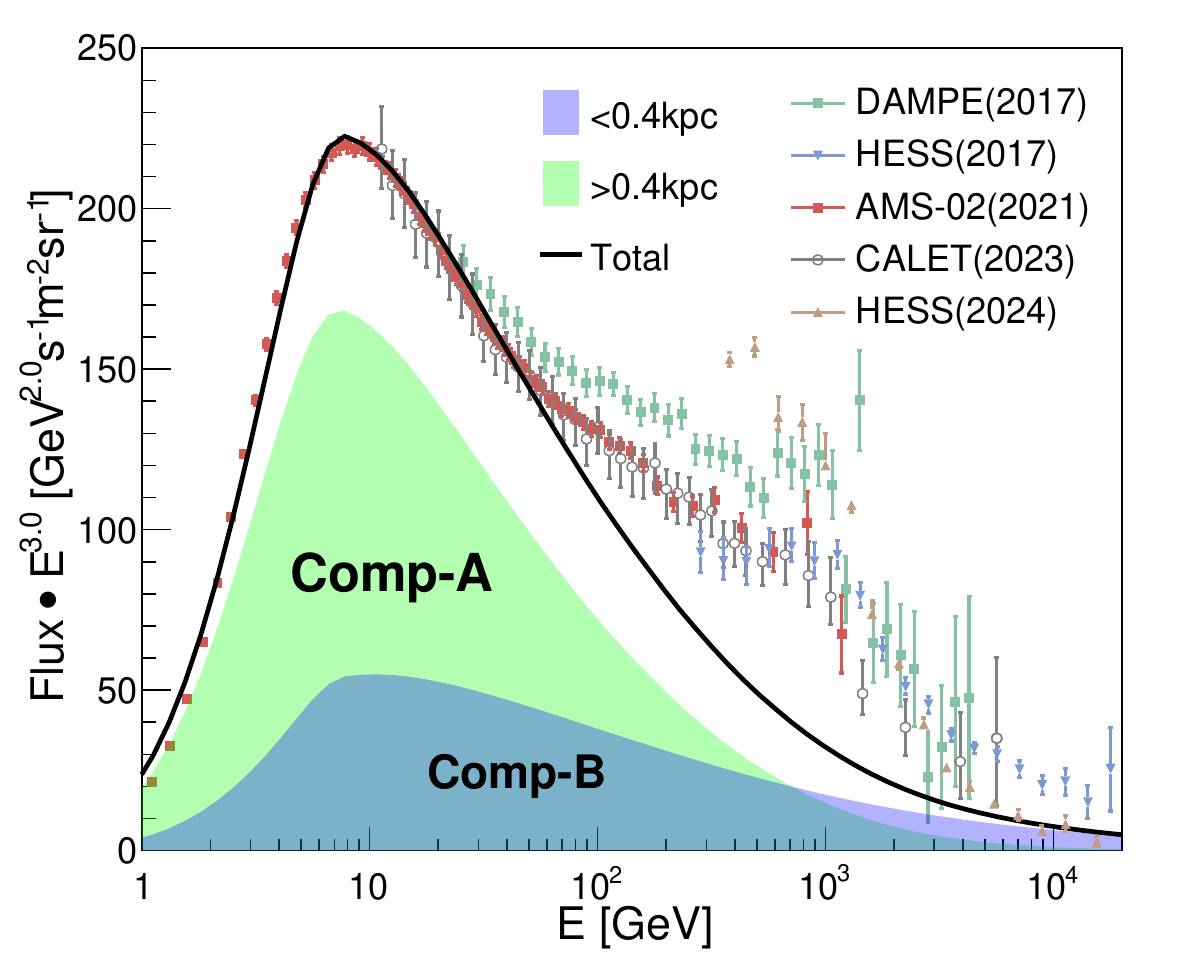}
    \end{subfigure}
    \caption{Partitioned calculation results of the SDP model and the traditional model compared with the observations from  AMS-02\cite{aguilar2021alpha}, DAMPE\cite{dampe2017direct}, HESS\cite{kerszberg2017cosmic,aharonian2024high}, and CALET\cite{adriani2023direct} experiments. The left panel displays the partitioned calculation results of SDP-2 within the SDP model, while the right panel presents the partitioned calculation results from the traditional model. The blue and green shaded areas represent the electron flux contributions from nearby sources $(R < 0.4 kpc)$ and distant sources $(R > 0.4 kpc)$, respectively. The black line indicates the total electron flux calculated by the models.}
    \label{fig:twosub}   
\end{figure*}

Applying the method mentioned above, we calculated the electron energy spectrum. The left panel of Figure 2 clearly illustrates the three-segment structure of the electron energy spectrum. The DAMPE, HESS, and CALET data employed represent the total electron data minus the positron data from AMS-02\cite{aguilar2021alpha}. The right panel of Figure 2 presents the measured results of the electron energy spectrum, along with the calculations from the traditional model and the SDP model. It is evident that the electron flux calculated using the SDP model accurately reproduces the complex structure of the observed electron energy spectrum, while the traditional model fails to capture the fine structure of the electron flux. SDP-1 and SDP-2 represent the results of the SDP model with parameter sets tailored for the DAMPE and AMS-02 experimental data, respectively.

To better understand the origins of the complex structure of the electron spectrum, we separately investigate the contributions of distant and nearby sources to the electron flux, referred to as Comp-A and Comp-B. As described in Sec.II , we use a standard of $0.4 kpc$, which is half the thickness of the inner diffusion halo, to distinguish between nearby and distant sources.
Figure 3 illustrates the results of the SDP-2 model and the partitioned calculations of the traditional model. In SDP model, Comp-A is concentrated in the low-energy range and exhibits distinct characteristics of cooling effects. In contrast, Comp-B is primarily focused in the high-energy range, displaying a clear transition from diffusion dominance to cooling dominance around $~700 GeV$.
However, the spectra of Comp-A and Comp-B in the traditional model are essentially consistent and cannot explain the complex structure of the electron energy spectrum. The partitioned calculation results provide strong support for the SDP model.

\section{CONCLUSIONS}

This work presents a clear and comprehensive propagation diagram that dynamically reveals the evolution mechanism of the electron energy spectrum. The dual diffusion timescale resulting from the SDP model’s double diffusion halo plays a crucial role in this evolution mechanism. Electrons originating from distant sources propagate in the outer halo, dominated by cooling effects, and exhibit results similar to those of traditional models. In contrast, electrons contributed by nearby sources are governed by diffusion and cooling, depending on their initial energy. Consequently, the three-segment power-law spectrum of electrons is produced by the outer halo cooling of distant source electrons, the inner halo diffusion of nearby source electrons, and the inner halo cooling of nearby source electrons. Based on the SDP model, we validate the aforementioned framework by separately investigating the contributions from nearby and distant sources. The results of model calculations successfully replicate the complex features of the electron energy spectrum and confirm the contributions of near-source and far-source electrons to the electron flux. The calculations presented provide support for the propagation diagram phenomenon we propose.

\begin{acknowledgments}
This work is supported by the National Natural Science
Foundation of China (No. 12275279, No. 12373105,
No. 12320101005).
\end{acknowledgments}

\nocite{*}

\bibliography{apssamp}

\end{document}